\documentclass[twocolumn,english,prl,amssymb,aps,superscriptaddress,showpacs,twocolumn,amsmath,showkeys,floatfix]{revtex4-2}
\usepackage[T1]{fontenc}
\usepackage[latin9]{inputenc}
\usepackage{geometry}
\usepackage[active]{srcltx}
\usepackage{textcomp}
\usepackage{amsmath}
\usepackage{graphicx}
\usepackage{color}
\usepackage{esint}
\usepackage{inputenc}
\usepackage{svg}
\makeatletter
\usepackage{babel}
\usepackage{physics}
\usepackage{hyperref} 
\makeatother

\begin{document}

\title{Multimode soliton collisions in graded-index optical fibers}

\author{Yifan Sun}
\affiliation{Department of Information Engineering, Electronics and Telecommunications, Sapienza University of Rome, Via Eudossiana 18, 00184 Rome, Italy}
\author{Mario Zitelli}
\affiliation{Department of Information Engineering, Electronics and Telecommunications, Sapienza University of Rome, Via Eudossiana 18, 00184 Rome, Italy}
\author{Mario Ferraro}
\affiliation{Department of Information Engineering, Electronics and Telecommunications, Sapienza University of Rome, Via Eudossiana 18, 00184 Rome, Italy}
\author{Fabio Mangini}
\affiliation{Department of Information Engineering, Electronics and Telecommunications, Sapienza University of Rome, Via Eudossiana 18, 00184 Rome, Italy}
\affiliation{Department of Information Engineering, University of Brescia, Via Branze 38, 25123 Brescia, Italy}
\author{Pedro Parra-Rivas}
\affiliation{Department of Information Engineering, Electronics and Telecommunications, Sapienza University of Rome, Via Eudossiana 18, 00184 Rome, Italy}
\author{Stefan Wabnitz}
\affiliation{Department of Information Engineering, Electronics and Telecommunications, Sapienza University of Rome, Via Eudossiana 18, 00184 Rome, Italy}


\begin{abstract} 
In this work, we unveil the unique complex dynamics of multimode soliton interactions in graded-index optical fibers through simulations and experiments. By generating two multimode solitons from the fission of an input femtosecond pulse, we examine the evolution of their Raman-induced red-shift when the input pulse energy grows larger. Remarkably, we find that the output red-shift of the trailing multimode soliton may be reduced, so that it accelerates until it collides with the leading multimode soliton. As a result of the inelastic collision, a significant energy transfer occurs between the two multimode solitons: the trailing soliton captures energy from the leading soliton, which ultimately enhances its red-shift, thus increasing temporal separation between the two multimode solitons.
\end{abstract}


\maketitle

\section{Introduction}

Solitons are nonlinear waves with particle-like behavior, with intriguing nonlinear dynamics. Solitons are ubiquitous in physics: they appear in different contexts, ranging from fluids
to plasmas \cite{Zabusky1965}, Bose-Einstein condensates \cite{Denschlag2000}, and nonlinear lattices \cite{Kartashov2011}.
In fiber optics, temporal solitons form due to a balance between nonlinear and dispersive effects. For ultrashort solitons (e.g., for durations <100 fs), the soliton spectrum becomes so broad, that the longer-wavelength spectral components experience Raman amplification, at the expense of shorter-wavelength components.
As a result, a continuous downshift of the mean frequency of the propagating soliton occurs \cite{Gordon1986}. This phenomenon is referred to as the Raman-induced soliton self-frequency shift (SSFS). 
The latter has been studied extensively \cite{Lee2008}, and many applications have been demonstrated, including wavelength-tunable pulse femtosecond sources \cite{Nishizawa1999}, analog-to-digital converters \cite{Xu2003}, and tunable delay lines \cite{Oda2006}.

As input pulse energy increases, higher-order solitons can be formed: these pulses have no binding energy, so that they are unstable against higher-order dispersion
and break-up into individual fundamental solitons \cite{Tai1988}.
The resulting fundamental solitons are subject to interaction forces: this is a problem of long-standing interest, thanks to the
richness of its associated physical effects \cite{Stegeman1999}. Soliton interactions have been extensively theoretically studied \cite{Wabnitz1995} in several different contexts \cite{Mitschke1987, Hause2009,Buch2016,Zhang2018,Balla2017,Balla2018}.
For example, soliton interactions play an important role in the formation of optical rogue waves \cite{Genty2010,Dudley2014,Kolpakov2016}, rogue solitons \cite{Armaroli2015}, and supercontinuum generation \cite{Herrmann2002}.

Manipulating soliton dynamics is a key challenge for many applications of solitons and solitary waves.
In particular, many efforts were made for controlling the SSFS,
e.g., for suppressing it by means of a negative dispersion slope \cite{Skryabin2003} and self-steepening \cite{Voronin2008}. 
On the other hand, SSFS can be enhanced by using
tapered fibers \cite{Bendahmane2013}, photonic crystal fibers \cite{Pant2010}, and metamaterials \cite{Xiang2011}. 
SSFS can also be controlled by using specially tailored Airy pulses \cite{Hu2015}. 
These studies mostly focus on the modification of dispersion and nonlinearity of a waveguide, because the relation between them affects the soliton properties, such as its temporal duration and power. Therefore, for a given waveguide [e.g., a singlemode fiber (SMF)], the most convenient and direct way to adjust the amount of SSFS ($\Omega_{\rm R}$) is to control the propagation length $z$ and the soliton power $P$.
The SSFS adjustment relation satisfies
$\Omega_{\rm R}\propto T_0^{-4}z\propto P^2  A_{\rm eff}^{-2}z$  , where $T_0$ is the soliton duration,
and $A_{\rm eff}$ is the transverse effective mode area of the SMF.


The advent of multimode fibers (MMFs) unlock the spatial degrees of freedom in nonlinear fiber optics \cite{Krupa2019}.  In recent years, various spatiotemporal nonlinear dynamical phenomena were intensively studied, mostly using graded-index (GRIN) MMFs. These include, for example, Kerr beam self-cleaning \cite{Krupa2017,Liu2016}, geometric parametric instability (GPI) \cite{Krupa2016}, spatial self-imaging \cite{Agrawal2019,Hansson2020}, spiral emission \cite{Mangini2021a}, multimode solitons (MMS) \cite{Renninger2013,Zitelli2021,Zitelli2021a}, spatiotemporal mode locking (STML) \cite{Wright2017}, and soliton molecules in MMF ML lasers \cite{Qin2018}.
{\color{red}}

In contrast to the case of SMFs, in GRIN fibers MMS with the same temporal shape can experience different 
nonlinear dynamics, depending on their modal composition: this is due to the fact that modes have different effective areas, hence nonlinear coefficients. 
As we shall see in this work, the multimode nature of MMS provides an additional flexibility in the control their SSFS. This can be achieved by properly managing the input modal composition of the MMS, which is something that cannot be done in their singlemode counterparts.

In our present study we unveil the previously undisclosed complex nonlinear dynamics of MMS interactions in GRIN optical fibers. Under appropriate input coupling conditions, one obtains that the fission of a femtosecond input pulse generates two separate MMSs. 
For relatively low input pulse energies, the SSFS of the two MMSs increases, as the input pulse energy grows larger.
Unexpectedly we found out that, above a threshold value of input energy, the trailing MMS reduces its rate of SSFS, in spite of the growing input pulse energy. As a result, a temporal collision with the leading MMS may occur: this collision is inelastic, which means that energy exchange between the solitons takes place. Specifically, after the collision the trailing soliton gains energy at the expense of the leading soliton: as a result, the SSFS (or group delay) of the trailing soliton grows larger, which leads to temporal separation among the two MMSs.

This paper is organized as follows. 
In section \ref{sec:model_simulation}, the model and physical parameters are introduced. Next, we study the properties of a soliton carrying only one specific mode in a GRIN fiber, by considering its group velocity (GV), group delay (GD) and SSFS. 
Next, the dynamics of two MMSs collision originating from a high-order MMS fission is theoretically studied. Finally, a detailed example of MMS temporal collision occurring during propagation along the GRIN fiber is discussed.
Section \ref{sec:experiments} introduces the experimental setup and results. We reproduce the predicted soliton collision features by varying the input pulse energy. 
Finally, Section \ref{sec:conclusion} draws the conclusions of the manuscript.

\section{Model and simulations}	\label{sec:model_simulation}		

\subsection{Model and parameters}			

The model we used to describe soliton propagation in GRIN MMF is based on the generalized multimode nonlinear Schr\"odinger equations (GMMNLSEs) \cite{Horak2012,Wright2018a}.
The field envelope in the fiber can be expanded on the basis of its eigenmodes:
\begin{equation}
\mathcal{E}(x,y,z,t) = \sum_{p=1}^{N}F_p(x,y)A_p(z,t),
\end{equation}
where $F_p(x,y)$ are the transverse mode patterns. 
For GRIN fibers, the eigenmodes are the Laguerre-Gauss (LG) modes.
These modes are orthogonal and normalized so that the mode amplitudes $A_p$ are expressed in $\rm \sqrt{W/m^2}$. The evolution of the field envelope $A_p$ of mode $p$ is governed by the GMMNLSEs \cite{Horak2012,Wright2018a}:
\begin{equation}
\frac{\partial A_p(z,t)}{\partial z}
=\mathcal{D}\{A_p(z,t)\}+\mathcal{N}\{A_p(z,t)\},
\label{eq:ggmnlse}
\end{equation}
where the dispersion terms are
\begin{equation}
\mathcal{D}=i(\beta_0^{(p)}-\beta_0^{(1)})A_p-(\beta_1^{(p)}-\beta_1^{(1)})\frac{\partial A_p}{\partial t}
+i\sum_{q\geq2}^4\frac{\beta_q^{(p)}}{q!}\left(i\frac{\partial}{\partial t}\right)^qA_p,
\label{eq:dispersion_terms}	
\end{equation}
and the nonlinear terms read as
\begin{equation}
\begin{aligned}
\mathcal{N}
&	= i\frac{n_2\omega_0}{c} \left(1+\frac{i}{\omega_0}\frac{\partial }{\partial t}\right) 	\sum_{l,m,n}^N [ (1-f_R)S_{plmn} A_l A_m A_n^*\\
&+f_R S_{plmn}A_l\int_{-\inf}^{t}\mathrm{d}\tau h_R*(A_m(z,t-\tau)A_n^*(z,t-\tau)) ] .	
\end{aligned}	
\label{eq:nonlinear_terms}
\end{equation}
The GRIN fiber we used in simulations and experiments has a $50$ $\rm \mu m$ diameter core with a parabolic refractive index profile, where the difference between the core center and the cladding is $\Delta n =0.015$. The field profile $F_p(x,y)$ for mode $p$ and corresponding $q$-th derivative of the propagation constant $\beta^{(p)}_q$ were directly calculated on the basis of the parabolic index distribution of the fiber \cite{Fallahkhair2008} [see details in Sec.~1 in Supplemental Document]. 
In the top panel of Fig.~\ref{fig:SSFS_single_mode_soliton}, we draw $F_p(x,y)$ for $p=1,\,\cdots,\,15$.
The nonlinear coupling coefficients of the modes are
\begin{equation}
S_{plmn}
= 	\frac{\int \mathrm{d}x \mathrm{d}yF_pF_lF_mF_n}
{\sqrt{\int \mathrm{d}x \mathrm{d}yF_p^2\int \mathrm{d}x \mathrm{d}yF_l^2\int \mathrm{d}x \mathrm{d}yF_m^2\int \mathrm{d}x \mathrm{d}yF_n^2}}.
\label{eq:raman_overlap_factors}
\end{equation}
For the nonlinear response of the fiber, we consider the standard parameters of a silica \cite{Stolen1989,Agrawal2013}: the nonlinear index $n_2 = 2.7 \times 10^{-20}\rm \,m^2 W^{-1}$, the coefficient of the Raman contribution to the Kerr effect $f_R=0.18$, the Raman response is $h_{\rm R}$, with the two time constants $\tau_1= 12.2\,\rm fs$  and $\tau_2= 32\, \rm fs$ \cite{Agrawal2013}.

For the simulations presented in the following sections, the fiber length is $2\rm \,m$. Dispersion coefficients with up to $N=4$ in Eq.~(\ref{eq:dispersion_terms}) are taken into account. The input pulses have $T_0=\rm 70\,fs$ duration, $\lambda_0=2\pi c/\omega_0=\rm 1400\,nm$ center wavelength, but may have different mode contents, as described in the following. 
\subsection{Group velocity of singlemode soliton in GRIN fiber.} \label{sec:features_MMS}



Let us start by considering the simplest case of a soliton carried by a single specific mode in the MMF. This case is helpful for understanding the mechanism of proper MMS collisions. Temporal collisions lead to a strong interaction of two solitons. A necessary condition for a collision to occur, is that the trailing soliton propagates faster than the leading one, so that the two solitons gradually approach each other. 

\begin{figure}[!ht]
\includegraphics[scale=1]{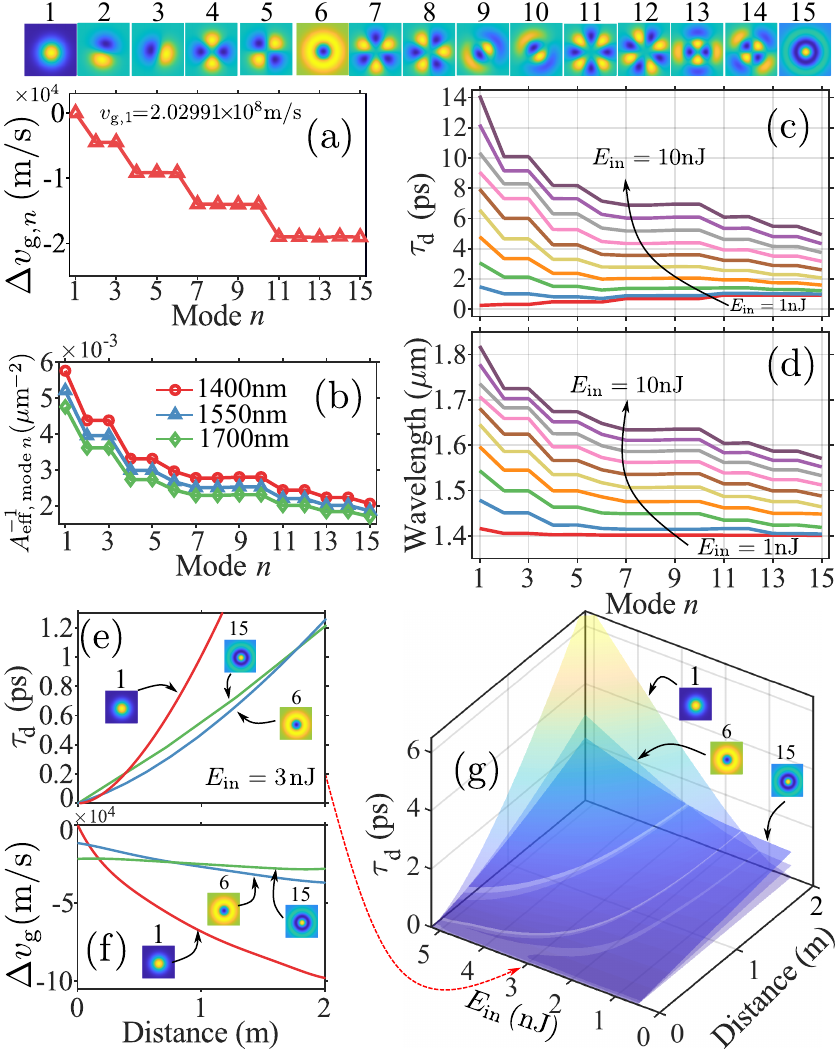}	
\centering
\caption{
Top panel: LG modes of the GRIN fiber used in this work. For compactness of notation, we use indexes to refer to these modes.
(a,b) Linear relative group velocity in (a) $\Delta v_{\mathrm{g},n}= (\beta_1^{(n)})^{-1}-(\beta_1^{(1)})^{-1}$  
and inverse of mode effective area  $A^{-1}_{\mathrm{eff,\, mode}\, n}$ in (b) vs. mode index $n$.
(c,d) Group delay and SSFS of singlemode solitons emerging from a 2 m long fiber, as a function of mode $n$.
(e,f) Group delay $\tau_{\rm d}$ and relative group velocity $\Delta v_{\mathrm{g}}$ of singlemode solitons with mode $n=1, \,6,\, 15$, as a function of propagation distance, when the input pulse energy is $E_{\rm in}=3\,\rm nJ$. 
(g) Evolution of group delay $\tau_{\rm d}$ of singlemode solitons (mode $n=1, \,6,\, 15$) vs. input energy $E_{\rm in}$ and propagation distance $z$. 
}
\label{fig:SSFS_single_mode_soliton}
\end{figure}

In the absence of nonlinearity, the GV of a propagating pulse mainly depends on the first-order dispersion coefficient: $v_{
\rm g,\it n}=(\beta_1^{(n)})^{-1}$. 
Since in GRIN fibers the $\beta_1^{(n)}$ values for the different modes are equally spaced, the relative GV $\Delta v_{\mathrm{g},n} = (\beta_1^{(n)})^{-1}-(\beta_1^{(1)})^{-1}\approx(\beta_1^{(1)}-\beta_1^{(n)})(\beta_1^{(1)})^{-2}$ are almost equally spaced, as shown in Fig.~\ref{fig:SSFS_single_mode_soliton}(a).
Therefore, in the absence of
nonlinear effects,
a pulse carried by a low-order mode (LOM) propagates faster than a pulse carried by high-order modes (HOMs). 
These different velocities can be easily identified by checking the group delay $\tau_{\rm d}$ [temporal shift of the pulse peak, with respect to a reference frame moving with the fundamental mode velocity $v_{\mathrm{F}} = 1/\beta_1^{(1)}=2.029\times10^8\rm\,m/s$]. By solving Eq.~(\ref{eq:ggmnlse}), we calculated the values of $\tau_{\rm d}$ for different singlemode pulses as a function of their mode index $n$, emerging from a $\rm2\,m$ long GRIN fiber. We carried out the calculation for different values of the input pulse energy: the corresponding results are shown in Fig.~\ref{fig:SSFS_single_mode_soliton}(c). Indeed, we can see that in the low input energy regime (e.g., for $E_{\rm in}=1\rm\,nJ$), LOMs have less GD than high-order ones.

However, as the input pulse energy increases, the GD of singlemode solitons is affected by the presence of (Raman) nonlinearity. 
We have checked the output pulses are not temporally broadened by dispersion when $E_{\rm in}\geq2\rm\,nJ$, which indicates that a soliton is formed, since the linear dispersion length of the fiber is $L_{\rm D}=T_0^2/|\beta_2|\rm=0.034\,m \ll 2\,m $.
In Fig.~\ref{fig:SSFS_single_mode_soliton}(c), we can see that the GD of LOM solitons increases faster with input energy, with respect to the case of HOM. Finally,  whenever $E_{\rm in}\geq3\rm\,nJ$, LOM solitons have a larger GD with respect to HOM. This means that, soliton GVs change with the input energy, so that by choosing a suitable input energy, two singlemode solitons could have the same GD at the fiber output. 

In the spectral domain, the SSFS of solitons exhibits a similar behavior to that of $\tau_{\rm d}$, as shown in Fig.~\ref{fig:SSFS_single_mode_soliton}(d). 
In fact, the mode distributions of the GD and the SSFS for different input pulse energies have a similar trend, see Fig.~\ref{fig:SSFS_single_mode_soliton}(c,d). This is because of the different effective mode areas $A_{\rm eff,\it n}$, which lead to different strengths of their nonlinearity. 
The strength of the Raman effect is proportional to the mode overlap factors $S_{plmn}$ [see Eq.~(\ref{eq:nonlinear_terms})]. 
For singlemode soliton in GRIN fibers, this term in Eq.~(\ref{eq:raman_overlap_factors}) is simplified, and it is equal to the inverse of the mode effective area $S_{nnnn}=A^{-1}_{\mathrm{eff,\, mode}\, n}$, which is shown in Fig.~\ref{fig:SSFS_single_mode_soliton}(b): 
the resulting decrease of the inverse effective area with mode $n$ is in agreement with the corresponding decrease of GD and SSFS, which are shown in Figs. \ref{fig:SSFS_single_mode_soliton}(b,d). 
In addition, the mode effective area increases with the wavelength [see Fig.~\ref{fig:SSFS_single_mode_soliton}(c)]. This indicates the nonlinearity strength reduces as the wavelength increases. 

\begin{figure*}[t]
\includegraphics[scale=1]{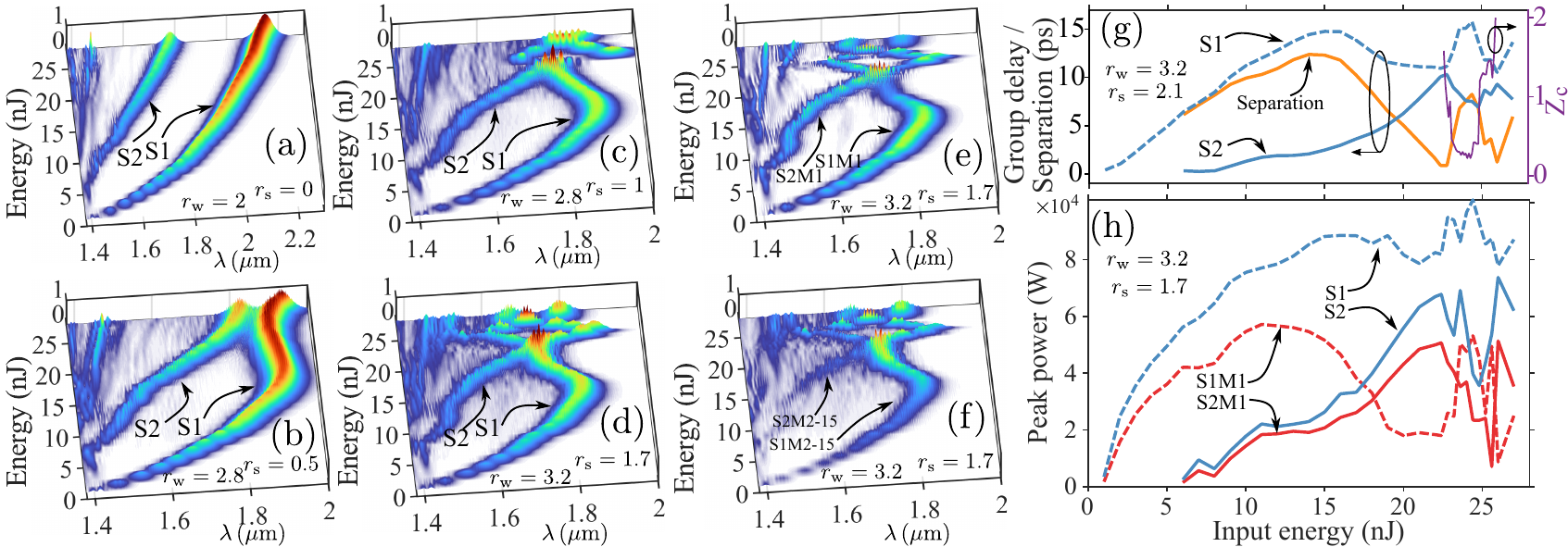}	
\centering	
\caption{
(a-d) Simulation results of output spectra vs. input pulse energy for an input Gaussian beam width $r_\mathrm{w}=2$ and center offset $r_\mathrm{s}=0$ in (a), $r_\mathrm{w}=2.8$ and $r_\mathrm{s}=0.5$ in (b), $r_\mathrm{w}=2.8$ and $r_\mathrm{s}=1$ in (c) and $r_\mathrm{w}=3.2$ and $r_\mathrm{s}=1.7$ in (d). The two MMSs are marked by S1 and S2. 
(e,f) Spectrum evolution of mode 1 (S1M1 and S2M1) and of the remaining 14 modes (S1M2-15 and S2M2-15) for the simulation in (d) are shown in panels (e) and (f), respectively. 
(g) Soliton GDs, their separation and the propagation distance $Z_c$ where the collision occurs for the simulation in (d), vs. input pulse energy.
(h) Peak power of S1 and S2 and their fundamental mode peak power (S1M1 and S2M1), vs. input pulse energy. 
}
\label{fig:soliton_vs_energy}
\end{figure*}

It is also interesting to consider how the velocity of singlemode solitons varies upon propagation along the GRIN MMF. One example of temporal delay $\tau_{\rm d}(z)$ of a singlemode soliton as a function of distance $z$ for modes 1, 6 , or 15, respectively, is shown in Fig.~\ref{fig:SSFS_single_mode_soliton}(e), for an input pulse energy $E_{\rm in}=3\,\rm nJ$. 
The local group velocity along distance $z$ can be calculated as
\begin{equation}
v_{\rm g}(z) = \frac{\dd z}{\dd t} = \left(\frac{\dd (t_{\rm frame}+\tau_{\rm d})}{\dd z}\right)^{-1}=\left(\beta_1^{(1)}+\frac{\dd \tau_{\rm d}}{\dd z}\right)^{-1}.
\end{equation}
This leads to the relative local group velocity with respect to a reference frame moving with the fundamental mode speed
\begin{equation}
\Delta v_{\rm g}(z) = v_{\rm g}(z)-(\beta_1^{(1)})^{-1}
\approx-(\beta_1^{(1)})^{-2}\frac{\dd \tau_{\rm d}}{\dd z}.
\label{eq:relative_GV}
\end{equation}
The values of $\Delta v_{\rm g}(z)$ for the three monomode solitons are shown in Fig.~\ref{fig:SSFS_single_mode_soliton}(f). As we can see in Fig.~\ref{fig:SSFS_single_mode_soliton}(f), initially (i.e., at $z=0$) all solitons have the GVs which are predicted according to Fig.~\ref{fig:SSFS_single_mode_soliton}(a). The GV of the soliton carried by mode $n=1$ is the largest in the beginning of the fiber but, due to the slowing down induced by the SSFS, it also experiences the fastest decay [see Fig.~\ref{fig:SSFS_single_mode_soliton}(f)]. This gives the soliton carried by mode $n=6$ (or mode $n=15$) the chance to catch up with the mode $n=1$ soliton at $\rm 0.2\,m$ (or $\rm 0.38\,m$) [see Fig.~\ref{fig:SSFS_single_mode_soliton}(e)]. The solitons carried by mode $n=6$ and $n=15$ also have the same GD at $\rm 1.73\,m$. 
These equal GD points vary when the input energy changes. The evolution of the GD $\tau_{\rm d}$ as a function of both input pulse energy $E_{\rm in}$ and propagation distance $z$ is shown in Fig.~\ref{fig:SSFS_single_mode_soliton}(g). Here the three white lines represent points of equal GDs for any pair singlemode solitons for a specific input pulse energy value. This indicates that two propagating solitons carried by different modes have the possibility to temporally overlap at a specific position in the fiber, owing to the nonlinear dependence of their GV.

For MMSs, the properties of the GV, GD and SSFS are more complex than in the case of singlemode solitons. 
For a MMS, the GD (under the influence of SSFS) not only depends on the values of peak
power, propagation length, and mode effective area, 
but it also varies with the specific mode composition. As discussed in details in Sec.~3 in Supplemental Document, MMSs carrying a larger portion of LOMs experience a larger amount of SSFS and GD with respect to MMSs carried by HOM.
Therefore, two MMSs with different mode compositions, will exhibit a different evolution of their GVs. As a result, under suitable conditions the two MMSs can acquire the same GD at a specific position in the fiber. 
For two MMSs originating from a MMS fission, these equal GD points provide the necessary condition for their collision, as we are going to see in the next section.


%
%
%
%
%

\subsection{Numerical simulations of soliton collision}	

Because of the previous considerations, we may expect that the fission of a high-order MMS could provide the testbed for the collision of two separate MMSs, carrying different mode contents. In this section we investigate evolution of MMSs. For doing that, we shall keep the same simulation parameters as before, except for increasing the input pulse energy, and varying the mode composition of the input pulse.


As a matter of fact, setting the appropriate input mode composition for the input pulse is a critical condition for controlling the occurrence of a soliton collision. 
In order to quantify the input mode content for a given input laser beam, we decomposed the input Gaussian beam with full-width-at-half-maximum (FWHM) $w$, and offset $s$ with respect to the fiber axis, on the basis of the LG modes.
These two parameters are normalized with respect to the FWHM $w_{\rm LG_{01}}=8.79$ $\rm \mu m$ of $|F_1(x,y)|^2$ at $\rm1400\, nm$.
Therefore, by tuning the dimensionless parameters $r_\mathrm{w}=w/w_{\rm LG_{01}}$ and $r_\mathrm{s}=s/w_{\rm LG_{01}}$, we can sweep over different modal compositions [Details about the dependence of the input mode content on these parameters can be found in Sec.~2 of the Supplemental Document]. 
Generally, the larger the beam size $w$, the higher the mode contents of the input beam.


\subsubsection{Output field evolution with input pulse energy}

We investigate by numerical simulations how the output field evolves with input pulse energy, as the input coupling conditions are varied. Figure~\ref{fig:soliton_vs_energy}(a) shows a first example of the input pulse energy dependence of the output spectra from a $\rm2\, m$ GRIN fiber. 
Here we consider injecting a beam with
$r_\mathrm{w} = 2$, $r_\mathrm{s} = 0$. 
As it can be seen, in the low energy regime ($E_{\rm in}\rm <6 \rm \,nJ$), only one MMS (S1) is formed. Whereas, a second MMS (S2) appears at $E_{\rm in}\rm >6 \rm \,nJ$, owing to the higher-order soliton fission. Both of these MMSs experience larger GD and SSFS, as the input energy increases [see \href{https://youtu.be/jyq5qdqNpB8}{\textit{\textbf{movie 1}}}]. As a result, the output wavelengths of S1 and S2 are $2.1$ $\rm \mu m$ and $1.75$ $\rm \mu m$ for $E_{\rm in}=25\,\rm nJ$, respectively.

By introducing a higher-order mode content at the fiber input, i.e., $r_\mathrm{w} = 2.8$, $r_\mathrm{s} = 0.5$, we obtain the result shown in Fig.~\ref{fig:soliton_vs_energy}(b). 
Here the second soliton S2 appears at the same input energy as before, i.e., $E_{\rm in}\rm >6 \rm \,nJ$. 
Unexpectedly, when $E_{\rm in}\rm > 17 \rm \,nJ$, the two MMSs exhibit a completely different dynamics, when compared with the one in Fig.~\ref{fig:soliton_vs_energy}(a): namely, now S1 undergoes a reduced amount of SSFS, as the input energy increases.
This brings the two MMSs both spectrally (and temporally) closer and closer to each other, however they remain spectrally distinct [see \href{https://youtu.be/0HnjynImtOA}{\textit{\textbf{movie 2}}}]. 
Whereas, for $E_{\rm in}\rm > 24 \rm \,nJ$ the two MMSs start to separate again.
In addition, the output wavelengths of S1 and S2 are $1.9$ $\rm \mu m$ and $1.8$ $\rm \mu m$ at $25\,\rm nJ$: the former is smaller than what previously reported in Fig.~\ref{fig:soliton_vs_energy}(a).

\begin{figure*}[t]
\includegraphics[scale=1]{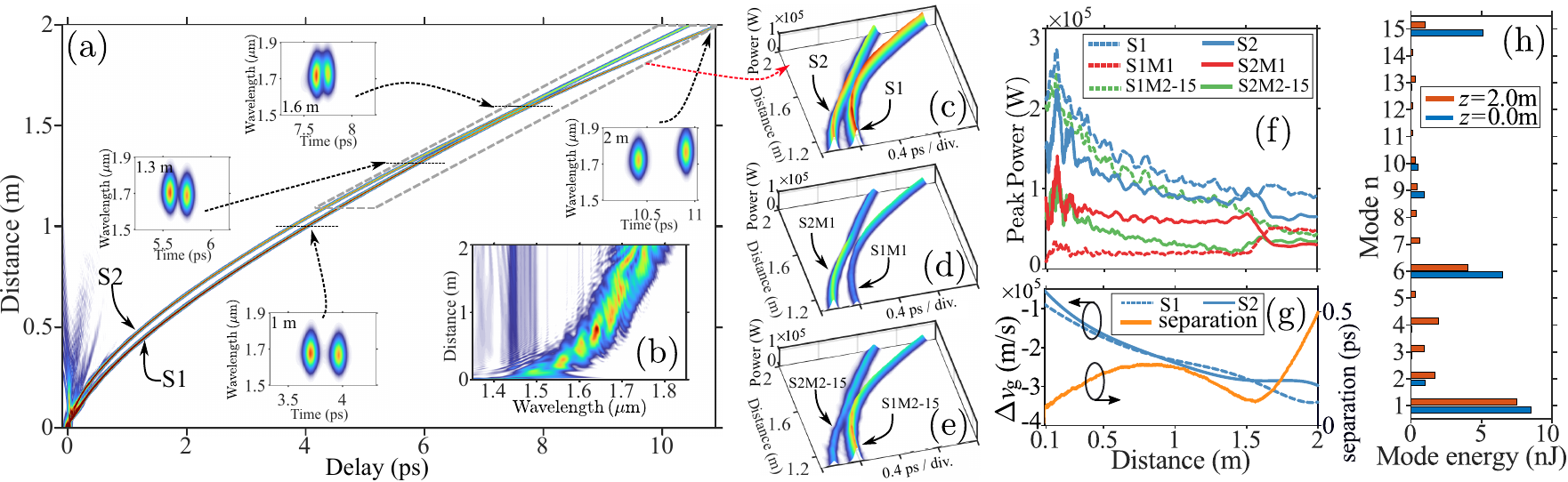}	
\centering
\caption{
One example of MMS evolution in the fiber in Fig.~\ref{fig:soliton_vs_energy}(d) ($r_\mathrm{w}=3.2$, $r_\mathrm{s}=1.7$), when input pulse energy is $E_{\rm in}\rm=22.7\,nJ$.
(a,b) Temporal and spectral evolution of field as a function of propagation length $z$, respectively.  Spectrograms at specific fiber lengths are inserted in (a).
(c-e) The temporal evolutions of total fields, mode 1 (M1), and sum of mode 2-15 (M2-15) correspond to the region of red dashed parallelogram in (a). 
(f,g) Evolution of total peak powers of solitons (S1, S2), their mode peak powers (S1M1, S2M1, S1M2-15, S2M2-15) in (f), and the relative group velocities of S1 and S2 and their temporal separation in (g) as a function of propagation length $z$.
(h)	Mode energy distribution comparison between the input and the output of the fiber.
}
\label{fig:soliton_vs_length}
\end{figure*}

By further increasing the HOM content at the fiber input, i.e., when setting $r_\mathrm{w} = 2.8$, $r_\mathrm{s} = 1$, we obtain the result shown in Fig.~\ref{fig:soliton_vs_energy}(c) [see temporal and spectral evolution of the two MMSs along the fiber in \href{https://youtu.be/tfUBz9g8mgM}{\textit{\textbf{movie 3}}}]. 
As can be seen, in this case for all input energies S1 experiences a much smaller SSFS when compared with the case of Fig.~\ref{fig:soliton_vs_energy}(b): as a result, S1 and S2 fully spectrally overlap at around $E_{\rm in}=\rm25\,nJ$.  
Upon further increasing the input energy, the spectra of the two MMSs separate, until they overlap again for $E_{\rm in}\rm=27\,nJ$. 
The two spectral fringe patterns at $E_{\rm in}\rm=25\,nJ$ and $E_{\rm in}\rm=27\,nJ$ imply that the two MMSs are very close to each other in the temporal domain.

By acting on the input mode composition, one may further tune the interactions between two MMSs. As an example, in Fig.~\ref{fig:soliton_vs_energy}(d), we report the case of $r_\mathrm{w} = 3.2$, $r_\mathrm{s} = 1.7$ [see their evolution along the fiber in \href{https://youtu.be/pnjt5DG8-Qg}{\textit{\textbf{movie 4}}}].
As one can visibly appreciate, in this case the dynamics is similar to that of Fig.~\ref{fig:soliton_vs_energy}(c); however, the collision wavelength and energy are reduced down to $1.72$ $\rm \mu m$ and $22.5\rm \,nJ$, respectively. 
Thus we can see that by increasing the input HOM content, one may shift the collision point to occur at shorter wavelengths and lower energies. This tendency is similar to the the behavior that we previously described for singlemode solitons in Fig.~\ref{fig:SSFS_single_mode_soliton}(d).

In order to reveal the physical mechanism leading to the unexpected evolution of the SSFS for S1, and the resulting soliton collisions, it is necessary to analyze how the mode content of the MMS evolves as the input energy is varied.
In order to do that, we decompose the total spectrum of Fig.~\ref{fig:soliton_vs_energy}(d) into its different modal components. In particular, this permits to highlight the specific contribution of mode 1. The latter is shown in Fig.~\ref{fig:soliton_vs_energy}(e), whereas the remaining mode content (ranging from mode 2 up to mode 15) is illustrated in Fig.~\ref{fig:soliton_vs_energy}(f). These spectra are labeled S1M1, S1M2-15 for S1 and S2M1, S2M2-15 for S2, respectively.
By comparing Figs.~\ref{fig:soliton_vs_energy}(e,f), we can ascribe the reduced SSFS of S1 to its decreased fundamental mode content S1M1. Indeed, the latter progressively reduces when increasing the input energy, i.e., at $>16\rm\,nJ$. To the contrary, the power increase in HOM content S1M2-15 does not provide a sufficient boost to the SSFS.

In order to quantitatively estimate the role of the modal content on the temporal evolution of MMSs, in Fig.~\ref{fig:soliton_vs_energy}(g) we plot the evolution with input pulse energy of the temporal GD (and temporal separation) of S1 and S2, respectively. Whereas in Fig.~\ref{fig:soliton_vs_energy}(h) we show the corresponding evolution of peak power of S1, S2, S1M1 and S2M1. 
As we can see, whenever $E_{\rm in}< 22.5$ nJ, the peak power of S1 and S2 increases as the input energy grows larger; for any input energy value, the power of S1 is larger than that of S2. 
However, the power of S1M1 starts to dramatically decrease at $E_{\rm}>15\,\rm nJ$. Correspondingly, the GD and SSFS experienced by S1 are both reduced, until they become comparable to the values of S2. This leads to generating temporally overlapping solitons when $E_{\rm}=22.5\,\rm nJ$, thus further confirming that the fundamental mode content plays a key role in determining the properties of the MMSs.

It is worth to mention that, in the input energy range between $\rm22.5\,nJ<\it E_{\rm in}\rm<26\,nJ$, the fiber position where the collision of two MMSs occurs depends on the specific input energy value.
Notably, for the specific cases of $E_{\rm in}=\rm22.5\,nJ$ or $E_{\rm in}=\rm26\,nJ$, the MMS collision occurs at $\rm 2\,m$. 
In Fig.~\ref{fig:soliton_vs_energy}(g) we plot the propagation distance $Z_c$ where the two MMSs have the closest temporal separation, i.e. the collision occurs, as a function of input pulse energy [see  \href{https://youtu.be/pnjt5DG8-Qg}{\textit{\textbf{movie 4}}}].
We found that the largest value of SSFS for S1 occurs when the input energy is $\rm24\,nJ$, which leads to MMS collision after $\rm0.3\,m$ of propagation. This is because, under this peculiar input condition, S1 gains energy from S2 after a minimal distance of propagation. Therefore, S1 propagates for the longest available fiber length, thus accumulating the maximum SSFS. For the other collision cases in this energy region, the smaller the SSFS (or GD) of S1, the longer the distance where collision occurs. One may expect that, for a longer fiber, collisions may occur for a wider energy range. 
A particular evolution of two colliding MMSs along the fiber is discussed in the next section. 

\subsubsection{Field evolution inside the GRIN fiber}

So far, we have shown the spectral features of the field which is observed at the fiber output. Thus, at this point, one may naturally wonder: how do MMSs exactly collide inside the fiber? 

An example of soliton collision in the fiber with the same parameters ($r_\mathrm{w} = 3.2$, $r_\mathrm{s} = 1.7$) of Fig.~\ref{fig:soliton_vs_energy}(d), and the input energy $E_{\rm in} \rm= 22.7\,nJ$, is depicted in Fig.~\ref{fig:soliton_vs_length}(a).
A 70 fs pulse is injected at the beginning of the fiber, and it splits into two MMSs, marked as S1 and S2. The GVs of both solitons are slower than the moving speed of the temporal reference frame. As a consequence, we can see that their temporal delay increases, as both solitons propagate along the fiber. 
In order to better display the temporal evolution of the solitons around the collision region [which occurs between $\rm1.1\,m$ and $\rm2\,m$, see the grey dashed parallelogram in Fig.~\ref{fig:soliton_vs_length}(a)], the total fields (S1,S2), the projection on mode $n=1$ (S1M1,S2M1) and the sum of the remaining modes (S1M2-15, S2M2-15) are processed by temporal translations, and re-plotted in Fig.~\ref{fig:soliton_vs_length}(c-e).
Here we can clearly see the power exchange between the two MMSs, as they approach each other, and collide around $\rm1.6\,m$. 

The evolution of the spectrum vs. propagation distance is shown in Fig.~\ref{fig:soliton_vs_length}(b). 
We may note the occurrence of an interference pattern in the spectral domain, which occurs in correspondence with the collision point.
The two interacting solitons can be better visualized by looking at spectrograms computed at $\rm 1\,m$, $\rm 1.3\,m$, $\rm 1.6\,m$, and $\rm 2\,m$, respectively, as reported in the insets of Fig.~\ref{fig:soliton_vs_length}(a).
We may notice that the two solitons have almost the same value of SSFS before the collision occurs. 
Conversely, after the collision, S1 acquires energy from S2, which leads to boosting both its SSFS and GD. 
The entire evolution of the temporal intensity profile (with corresponding spectrograms) along the fiber is shown in \href{https://youtu.be/_OSp7mxS1as}{\textit{\textbf{movie 5}}}.

In order to better display the influence of the mode content on the evolution of a MMS, 
in Fig.~\ref{fig:soliton_vs_length}(f) we plot the peak power of both S1 and S2, along with their fundamental (S1M1, S1M1) and HOM content (S1M2-15, S2M2-15), as a function of the propagation distance. 
Furthermore, by extracting from Fig.~\ref{fig:soliton_vs_length}(a) the soliton delay $\tau_{\rm d}(z)$ as a function of propagation distance $z$, we calculated 
the relative GV $\Delta v_{\mathrm{g}}(z)$ by using Eq.~(\ref{eq:relative_GV}), as well as the temporal separation between the two solitons, both which are plotted in Fig.~\ref{fig:soliton_vs_length}(g).

Based on Figure~\ref{fig:soliton_vs_length}(f,g), we may highlight the following three main phases of the collision process: 
(i) \emph{Before the collision.} After the fission which takes place at $\rm 0.1\,m$, the input pulse is splitted into two separate MMSs. As a result, S2 is generated, which initially propagates faster than S1 [see Fig.~\ref{fig:soliton_vs_length}(g)]. The peak power of both S1 and S2 exhibits an asynchronous oscillatory behavior [see Fig.~\ref{fig:soliton_vs_length}(f)].
As discussed before, the SSFS and the GD are influenced by the Raman effect, whose impact is dominated by the contribution of the fundamental mode. Therefore, although the total power of S2 is smaller than that of S1, the peak power of S2M1 is larger than that of S1M1. Thus, S2 experiences a larger Raman effect than S1. Hence, we see that the GV of S2 reduces faster than that of S1, until it gets even smaller that the velocity of S1 at $\rm0.86 \,m$. (ii) \emph{At the Collision.} As a result of the GV dynamics, the two MMSs eventually collide at $\rm1.6\, m$. 
It is worth mentioning that the occurrence of a collision can be fully ascribed to the multimode nature of S1 and S2. 
As a matter of fact, when considering soliton fission in singlemode fiber, the trailing soliton S1 with a larger energy can never accelerate and reach S2, which removes the possibility of any soliton collision to occur. 
Due to the inelastic collision, S1 gains energy from S2. Specifically, the energy of S1M1 increases, while S2M1 decreases. 
(iii) \emph{After the collision}. Thanks to its increased fundamental mode content, now S1 undergoes a larger SSFS, and a lower GV with respect to S2. Therefore, the two MMSs progressively separate in time, without experiencing any further interaction. 
Finally, in Fig.~\ref{fig:soliton_vs_length}(h) we compare the mode content at the fiber input and output. Here, we can see that the radial modes (modes with $n=1, \,6, \, 15$) lose their energy, which is conversely acquired by non-radial modes.


\section{Experiments}	\label{sec:experiments}

In order to confirm the simulation results in Sec.~\ref{sec:model_simulation},
we have carried out a set of experimental tests. 
Let us start by describing the experimental setup, before reporting the observations which closely match our theoretical predictions.

\subsection{Experimental setup}

\begin{figure}[!h]
\includegraphics[scale=1]{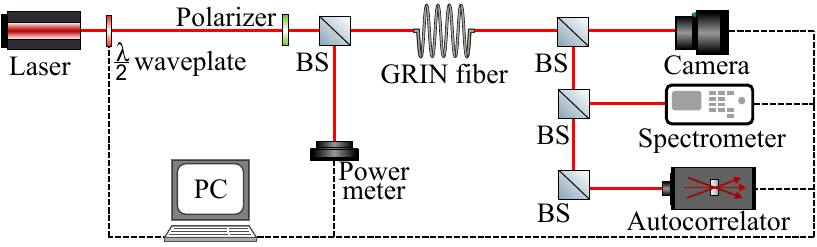}	
\centering	
\caption{ 
Sketch of the experimental set-up. 
}
\label{fig:experiment_setup}
\end{figure}

In our experiments, we used the same type of GRIN fibers as previously described.
The experimental setup is shown in Fig.~\ref{fig:experiment_setup}. 
Linearly polarized optical pulses (with $\rm 70\,fs$ temporal duration, $\rm 1400\,nm$ center wavelength and $\rm 100\,kHz$ repetition rate) are emitted by a hybrid optical parametric amplifier (Lightconversion ORPHEUS-F), pumped by a femtosecond Yb-based laser (Lightconversion PHAROS-SP-HP). 
The input pulse energy is controlled by rotating the computer motorized $\lambda/2$ waveplate, as shown in Fig.\ref{fig:experiment_setup}. A beam splitter (BS) is used for monitoring the input power by means of a power meter (Thorlabs PM16-122). The laser beam, which has a Gaussian profile ($M^2 = 1.1$), is injected by means of a 50 mm lens into the GRIN fiber, with a diameter of approximately 30 $\rm \mu m$ at $1/e^2$ of peak intensity on the fiber input facet.
At the fiber output, the beam is collected by an achromatic microlens, and separated into three paths, in order to measure the output beams near field, their spectra, as well as the MMS temporal separation, by using an InGaAs camera (Hamamatsu C12741-03), a spectrometer (Fastlite Mozza), and an autocorrelator (APE pulse check 50), respectively. 
The latter is appropriately equipped with a $\rm1500\,nm$ longpass filter, in order to extract Raman soliton out of the total output spectrum.

The accurate control of the coupling conditions of the input pulses into the GRIN fiber is a critical condition for studying the dynamics of MMS collisions.
Whenever the input beam is symmetrically coupled at the center of the fiber, one obtains a spectral evolution which is similar to the case reported in the simulation of Fig.~\ref{fig:soliton_vs_length}(a).
Therefore, in order to unveil the peculiar MMS collision-induced spectral dynamics, we offset the input laser beam by around $7$ $\rm \mu m$ with respect to the center of the fiber core. This leads to the generation of MMSs with far greater HOM content at the beginning of the fiber. This beam offset and the value of the input pulse energy need to be finely tuned, until we may find out the occurrence of an interference fringe spectrum. As we have seen before, this is a signature of the collision of two MMSs at the fiber output. Once this is done, the only parameter to be adjusted is the input pulse energy, in order to record the corresponding nonlinear evolution of the output spectrum.


\subsection{Experimental results}

Since we aim at experimentally retrieving the simulation predictions of Sec.~\ref{sec:model_simulation}, we used $L_1=\rm2\,m$ of GRIN fiber. Next, we further confirmed the generality of our findings by using a $L_2=\rm10\,m$ long GRIN fiber span. 

\begin{figure}[t]
\includegraphics[scale=1]{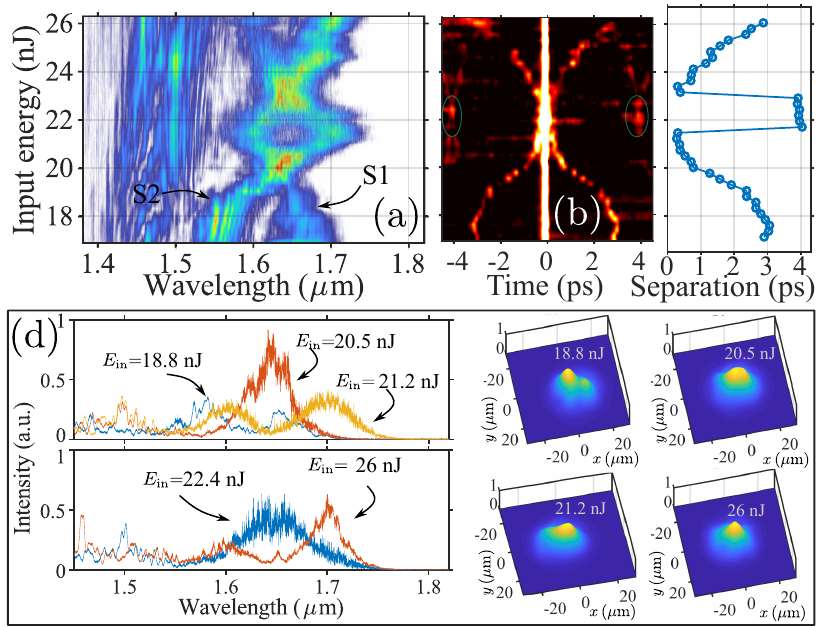}	
\centering	
\caption{Experimental results for a $\rm2\,m$ fiber. (a-c) Spectral evolution in (a), autocorrelation in (b), and temporal separation between the solitons vs. input energy. (d) Selection of four spectra with their corresponding output beams, for different input pulse energies as in (a). 
}
\label{fig:experiment_2m}
\end{figure}	

In Fig.~\ref{fig:experiment_2m} we show the results of our experiments with the $\rm2\,m$ fiber. Specifically, in Figs.~\ref{fig:experiment_2m}(a,b) we illustrate the measured output spectra, along with their corresponding autocorrelation traces, at different input pulse energies. 
For $E_{\rm in}< 20\rm \,nJ$, we can clearly identify the presence of two distinct solitons, both in the spectral domain and in the temporal domain. The temporal separation of the two solitons can be inferred by the autocorrelation traces in Fig.~\ref{fig:experiment_2m}(c). Fig.~\ref{fig:experiment_2m}(a) shows that, when $E_{\rm in}$ increases, the spectra of the two MMSs get progressively closer in the frequency domain. 
However, one cannot clearly distinguish the presence of two separate solitons by just examining the fringe pattern which appears in the spectrum, for energies between $20\rm\, nJ<E_{\rm in}< 21\rm \,nJ$. This is why we need to complement our spectral measurements with the temporal domain results of Figs.~\ref{fig:experiment_2m}(b,c), which reveal the presence of two MMSs with a separation of less than $\rm0.5\,ps$ for $E_{\rm in}= 20.5\rm \,nJ$.
For $21\rm\, nJ<E_{\rm in}< 23\rm \,nJ$, we detected the presence of a single dominant soliton in the middle of autocorrelation trace of Fig.~\ref{fig:experiment_2m}(b). Nevertheless, this figure also shows the occurrence of a weak peak at $\rm \pm4\,ps$ for input energies in the same range $21\rm\, nJ<E_{\rm in}< 23\rm \,nJ$. This indicates that the two solitons do not overlap in the temporal domain at the fiber output. To the contrary, they are well separated in time after the occurrence of a collision at a previous position $Z_c<L_1$ in the fiber. The presence of well-separated solitons are confirmed by the occurrence of distinct spectra in Fig.~\ref{fig:experiment_2m}(a) in the same input energy range. The two solitons are separated at the fiber output  because the trailing soliton S1 acquires energy from the leading soliton S2 at the collision point in the fiber. This leads to enhancing the red shift for S1. For input energies larger than the collision region (i.e., for $E_{\rm in}> 23\rm \,nJ$), Figs.~\ref{fig:experiment_2m}(b,c) shows that the two MMSs are again well separated at the fiber output. 
In addition, we may note in Fig.~\ref{fig:experiment_2m}(d), showing four beams and their corresponding spectra, that different MMSs enhibit a multimode transverse profile which varies with the input pulse energy. 

\begin{figure}[t]
\includegraphics[scale=1]{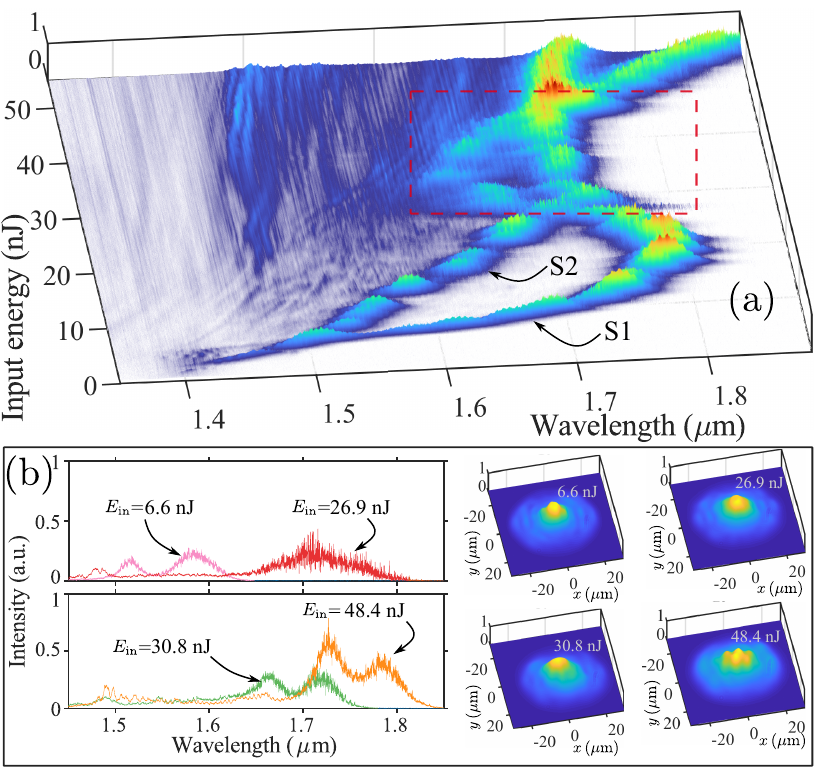}	
\centering	
\caption{Experimental results with a $\rm10\,m$ long GRIN fiber. (a) Spectral evolution vs. input energy. (b) Selection of four spectra  and output beams, for different input energies in (a). 
}
\label{fig:experiment_10m}
\end{figure}

Now, it is interesting to both qualitatively and quantitatively compare simulation and experimental results. Fig.~\ref{fig:experiment_2m}(d) shows that the first soliton overlap point at the fiber output occurs at $1.65$ $\rm \mu m$, for an input energy of $20.5\rm \,nJ$.
These values are slightly lower than the soliton overlap point that found numerically at the soliton overlap point, which occurs for $\rm22.5\,nJ$ [cfr. Fig.~\ref{fig:soliton_vs_energy}(d)]. The small discrepancy is likely to be due to the limited number of modes which is used in simulations.

So far, we have shown spectral and temporal properties of the field emerging at the fiber output. In order to experimentally monitor the collision dynamics along the fiber, similarly to what reported by simulations in Sec.~\ref{sec:model_simulation}, one would need to carry out a cut-back experiment. However, this is a challenging task, since performing a cut-back experiments may lead to changing the bending properties of the fiber, which can be detrimental for our study. Therefore, we limit ourselves to validate our conclusions by repeating the experiments with a longer, $\rm10\,m$ span of GRIN fibers.
In Fig.~\ref{fig:experiment_10m}(a) we show the corresponding measured output spectrum, again as a function of the input energy $E_{\rm in}$. 
As we can see, the solitons experience larger amounts of SSFS, when compared with the former result in Fig.~\ref{fig:experiment_2m}(a).
In a first stage (i.e., for $E_{\rm in}< 6\rm \,nJ$) a MMS (S1) is formed, whose wavelength increases with $E_{\rm in}$.
For $E_{\rm in}> 6\rm \,nJ$, the fission of the input pulse generates an additional MMS (S2). 
When $E_{\rm in}$ grows larger, S1 and S2 undergo different amounts of SSFS. Again, when $E_{\rm in}>20\rm  \rm \,nJ$, we observed that the SSFS of S1 is reduced as the input pulse energy increases: this appears as a relative ``blue-shift'' in Fig.~\ref{fig:experiment_10m}(b). Once again, the observed spectral evolution is qualitatively remarkably similar to simulation predictions. 
A spectral overlap of the two solitons is reached for $E_{\rm in}\simeq 26.9\rm \,nJ$.
The remarkable output spectra, corresponding to input energies such that collision occurs at some point inside the fiber, are marked by the red dashed box in Fig.~\ref{fig:experiment_10m}(b).
Finally, for $E_{\rm in}> 45\rm \,nJ$, the two solitons S1 and S2 clearly separate again. Examples of four spectra with their corresponding beams at different $E_{\rm in}$ are illustrated in Fig.~\ref{fig:experiment_10m}(b).
The detailed evolution of spectra can be seen in \href{https://youtu.be/g7jRTYVUZK8}{\textit{\textbf{movie 6}}}. 

\section{Conclusions}	\label{sec:conclusion}

To summarize, in this work we have numerically and experimentally studied the interaction of MMSs, resulting from the fission of femtosecond pulses in GRIN fibers. We have revealed the surprising result that, as a result of the variation of the MMS mode content, the SSFS of the trailing MMS exhibits is reduced, in spite of the growing energy of the input pulse. 
This is an anomalous behavior, which has no counterpart in the realm of singlemode fiber solitons.
The physical mechanism behind such behavior is the variation, with input energy, of the mode composition of the MMS that results from the fission of the input pulse. Specifically, the fundamental mode is depleted in favour of HOMs. This results in an input energy dependence of the group velocity of the trailing MMS. As a result, an inelastic collision may occur between the two fission-generated multimode solitons. In turn, the collision leads to a redistribution of both energy and mode content between the two interacting solitons. 
The nonlinear collision dynamics predicted by numerical simulations is well confirmed by experiments. 
From a fundamental standpoint, our analysis unveils the previously undisclosed complexity of MMS interactions. 
In addition, our results deepen the current understanding of the dynamics of MMSs, which may lead to rogue wave formation, supercontinuum generation, and spatiotemporal mode-locking in multimode fiber lasers.

\section*{Acknowledgments}
This work was supported by European Research Council (740355), Marie Sklodowska-Curie Actions (101023717), Ministero dell'Istruzione, dell'Università e della Ricerca (R18SPB8227), and Sapienza Università di Roma (AR22117A8AFEF609, AR22117A7B01A2EB).

\bibliography{soliton_collision_references}


\section{Supplementary Materials}

\subsection{Fiber dispersion and mode effective area.}\label{sec:fiber_dispersion}

This section provides with more details of the eigenmodes of the GRIN fiber considered in this work. Specifically, we focus on their dispersion relation as well as their effective mode areas.

As mentioned in the main text, we used the GRIN fiber with a parabolic refractive index profile, having a core diameter $R=50$ $\rm \mu m$ and whose cladding is made of undoped silica. Therefore, the Sellmeier equation is used to generate frequency dependent refractive index of the cladding $n_{\rm cl}(\omega)$. 
Being the refractive index difference between the core center and the cladding $\Delta n = n_{\rm co}-n_{\rm cl} =0.015$, the refractive index of the whole fiber can be written as
\begin{equation}
	n(x,y,\omega) =\left\{
	\begin{array}{lll}
		&n_{\rm cl}(\omega)+ \Delta n\cdot \left(1- \frac{x^2+y^2}{R^2}\right), &x^2+y^2<R^2\\
		&n_{\rm cl}(\omega), &x^2+y^2>R^2
	\end{array}
	\right .
\end{equation}
Based on the parabolic profile of core refractive index
, the first 15 transverse eigenmodes $F(x,y,\omega_s)$ and their corresponding eigenvalues (modal effective indices) $n_p(\omega_s)$ are calculated at chosen frequencies $\omega_s$ by the semivectorial finite difference method \cite{Fallahkhair2008}. 
These mode patterns are shown in top panel of Fig.~\ref{fig:mode decomposition}. 
The propagation constants $\beta^{(p)}(\omega_s)$ can be calculated as
\begin{equation}
	\beta^{(p)}(\omega_s)=\frac{\omega_sn_p(\omega_s)}{c},
\end{equation}
where $c$ is the light velocity in vacuum.

\begin{figure*}[tbp]
	\centering
	\includegraphics[scale=1]{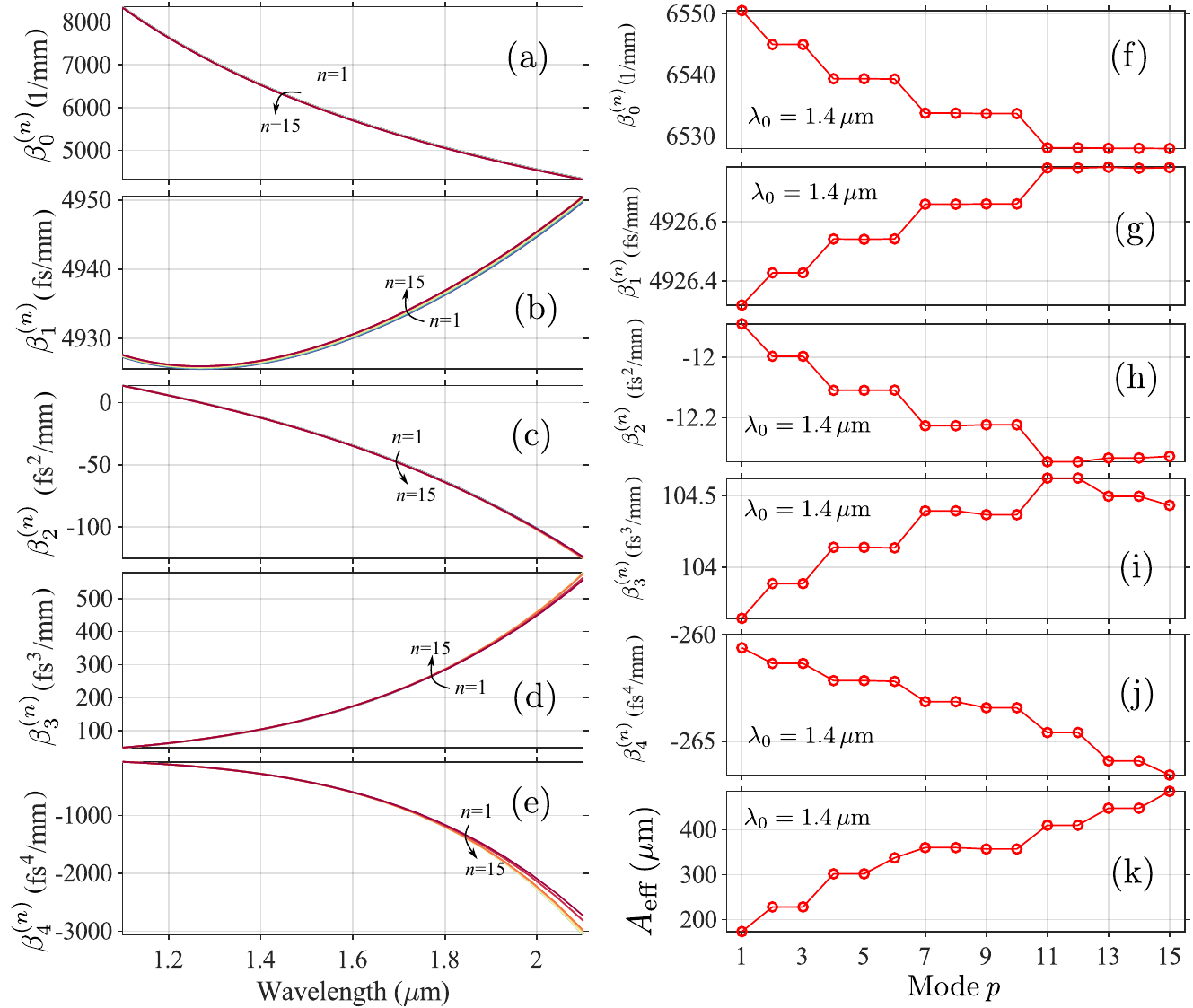}	
	\caption{Dispersion relations and effective areas of GRIN fiber eigenmodes. (a-e) First 5 order dispersion coefficients $\beta^{(p)}_n(\omega)={\partial^n(\omega) \beta^{(p)}}/{\partial \omega^n}$ as a function of the wavelength. (f-j) First 5 order dispersion coefficients 
		at $\lambda = \rm1400\,nm$ as a function of mode $p$. (k) Effective mode area at $\lambda = 1400\rm nm$ vs mode $p$.}
	\label{fig:fiber_dispersion}
\end{figure*}

By using a finite difference method, the $n$-th order derivative of $\beta^{(p)}(\omega)$ with respect to $\omega$, i.e., 
\begin{equation}
	\beta_n^{(p)}(\omega)=\frac{\partial^n \beta^{(p)}}{\partial \omega^n}.
\end{equation}	  
can be easily computed. The ensemble of $\beta_n^{(p)}$ are referred to as dispersion parameters. 
The results of the first five order derivatives are 
shown in Fig.~\ref{fig:fiber_dispersion} (a-e). As it can be seen, the dispersion relation at each order is rather similar for all of the modes.
Nonetheless, one may notice that for even orders [see Fig.~\ref{fig:fiber_dispersion} (a,c,e)], the dispersion parameters decrease as the wavelength grows larger. Moreover, dispersion turns out to increase within the mode $p$. This behavior is reversed for odd orders of dispersion [see Fig.~\ref{fig:fiber_dispersion} (b,d)]. Indeed, the dispersion parameters quench whenever either the wavelength or the mode index increase.

In order to go beyond the finite difference approximation and thus obtaining a more accurate evaluation of $\beta_n^{(p)}(\omega)$, we made a polynomial fit of the ensemble of $\beta^{(p)}(\omega_s)$. This allows for calculating the values of $\beta_n^{(p)}(\omega)$ at a chosen frequency $\omega_0$ as 
\begin{equation}
	\beta_n^{(p)}=\frac{\partial^n \beta^{(p)}_{\rm fit}}{\partial \omega^n}\bigg|_{\omega = \omega_0},
\end{equation}	
where $\beta^{(p)}_{\rm fit}$ is the fitting function. As mentioned in the main text, we choose $\omega_0=2\pi c/\lambda_0$ and $\lambda_0=1400\rm\,nm$. The dispersion parameters obtained by such fitting based method which are shown in Fig.~\ref{fig:fiber_dispersion} (f-j) are the ones that are used in the simulations reported in the main text based on the GMMNLSEs in Eq.~(1). 

One may notice a small deviation of $\beta_3^{(n)}$ for the highest-order modes [i.e. mode 13, 14, 15] in Fig.~\ref{fig:fiber_dispersion}. This is because high-order modes with large area are more influenced by the truncation of parabolic refractive index. Indeed, as wavelength increases, this distortion becomes larger, as it can be directly seen in Fig.~\ref{fig:fiber_dispersion} (e).
Hence, we recall the definition of mode effective area as 
\begin{equation}
	A_{\rm eff,\,mode\,\it p} = \frac{\left(\int\int|F_p(x,y)|\dd x\dd y\right)^2}{\int\int|F_p(x,y)|^4\dd x\dd y},
\end{equation}	
which is equal to reciprocal of the mode overlap coefficient $A_{\mathrm{eff,\, mode}\, p}=S_{pppp}^{-1}$ in Eq.~(4) in the main text
. 
An example of $A_{\rm eff}$ at $\rm1400\,\rm nm$ wavelength is 
plotted in Fig.~\ref{fig:fiber_dispersion}(k). 

%
%
%
%
%
%
%
%

\pagebreak
\subsection{Mode decomposition of a input Gaussian beam.} \label{sup:mode_decomposition}

In this section, we present the method used for computing 
the input mode distribution associated to the experimental laser-fiber coupling conditions. 


It is convenient to expand the field associated to the propagation of an optical beam in a multimode fiber on its eigenmodes basis $F_n(x,y)$: 
\begin{equation}
	\mathcal{E}(x,y,z,t) = \sum_{n=1}^{N}F_n(x,y)A_n(z,t),
\end{equation}
where the mode profiles are both normalized 
and 
mutually orthogonal, i.e.,
\begin{equation}
	\int_{-\infty}^{\infty}\int_{-\infty}^{\infty}F_nF_m\dd{x}\dd{y} = \delta_{nm}.
\end{equation}
$A_n(z,t)$ is the time dependent amplitude of the $n$-th mode at the propagation distance $z$, which appears in the GMMNLSEs in Eq.~(1) of the main text. Being the mode field power equal to $|A_n(z,t)|^2$, the input pulse energy reads as
\begin{equation}
	E_{\rm in} = \sum_{n=1}^{N}\int_{-\infty}^{\infty}|A_n(z=0,t)|^2\dd t.
\end{equation}

Here, we suppose that the temporal shape of all of the modes is the same at the input ($z=0$). Specifically, we consider a Gaussian shape which is centered at $t_0$ and having a duration $T_{\rm FHWM}=T_0/1.665$, i.e.,
\begin{equation}
	A_n(z=0,t) = C_n \sqrt{\frac{E_{\rm in}}{T_0\sqrt{\pi}}}e^{-\frac{(t-t_0)^2}{2T_0^2}},
\end{equation}
where the parameter $|C_n|^2$ is the power fraction associated to the mode $n$, that is also referred to as mode content or mode coefficients. The coefficients $|C_n|^2$ are normalized to the total power, so that they satisfy
\begin{equation}
	\sum_{n=1}^{N}|C_n|^2=1.
\end{equation}

Let us now focus on the spatial features. As described in the main text, we consider the case of a Gaussian beam which is injected into the fiber core with an offset ($x_0,y_0$) with respect to the fiber axis. Therefore, the field at the fiber input facet can be written as 
\begin{equation}
	\mathcal{E}_{\rm in}(x,y) = \frac{1}{\sigma\sqrt{\pi}}e^{-\frac{(x-x_0)^2+(y-y_0)^2}{2\sigma^2}},
	\label{eq:input_beam}
\end{equation}
where $w=\sigma/1.665$ is the FWHM of $|\mathcal{E}_{\rm in}(x,y)|^2$, which satisfies
\begin{equation}
	\int_{-\infty}^{\infty}\int_{-\infty}^{\infty}|\mathcal{E}_{\rm in}(x,y)|^2\dd x\dd y = 1.
\end{equation}
We can obtain the mode coefficients $C_n$ by projecting Eq.~(\ref{eq:input_beam}) onto the eigenmodes basis 
\begin{equation}
	C_n = \int_{-\infty}^{\infty}\int_{-\infty}^{\infty}\mathcal{E}_{\rm in}(x,y)\cdot F_n(x,y)\dd x\dd y.
\end{equation}

\begin{figure*}[!ht]
	\centering
	\includegraphics[scale=1]{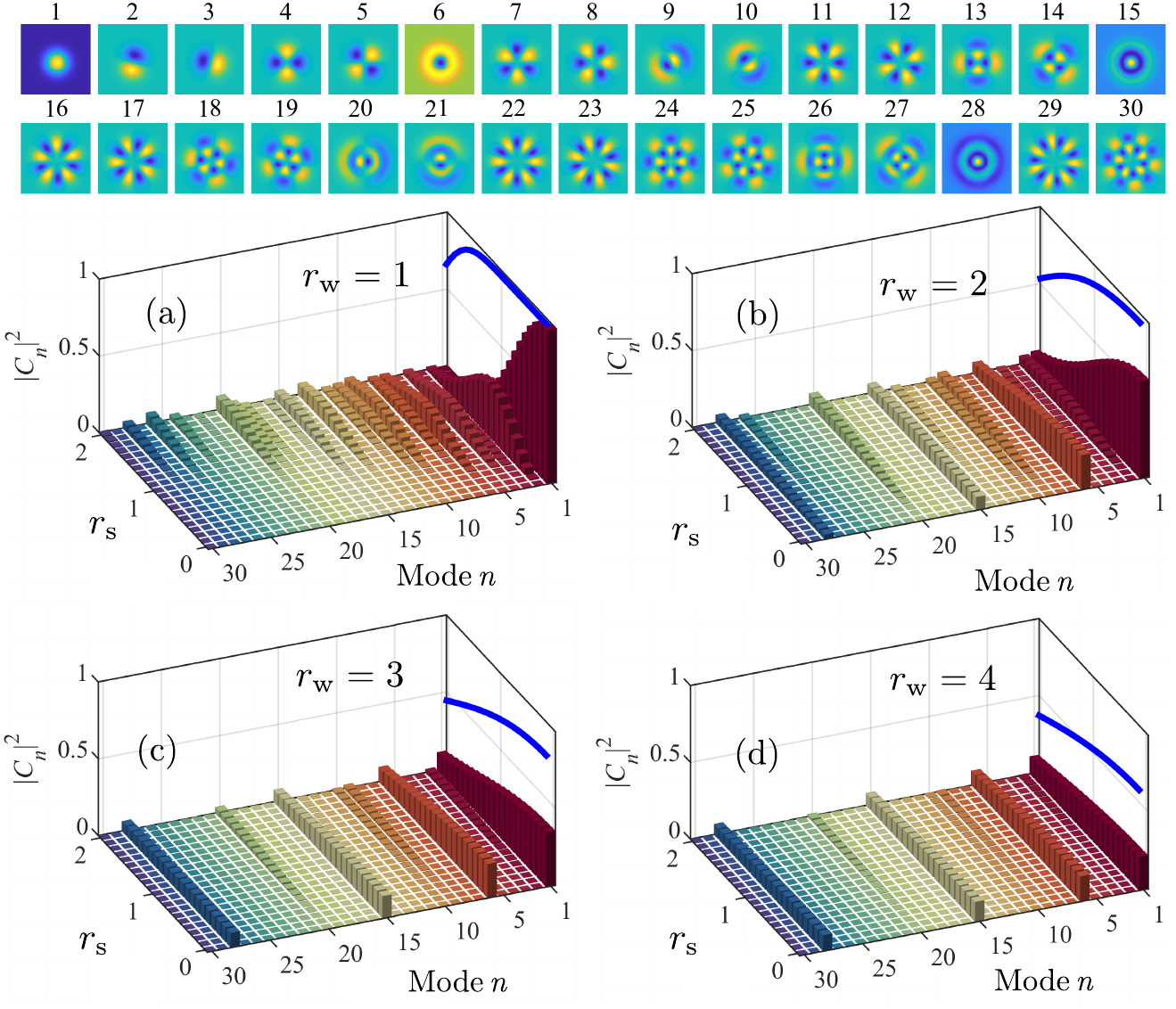}	
	\caption{Mode decomposition of a Gaussian beam into the first $N=30$ modes of a GRIN fiber (shown on the top of the figure). (a-d) Computed mode coefficients $|C_n|^2$ as a function of the beam offset with respect to the fiber axis ($r_{\rm s}$) for different beam sizes $r_{\rm w}=1,\,2,\,3,\,4$. The blue lines in (a-d) represent the value of $\sum_{n=1}^{N}|C_n|^2$ vs. $r_{\rm s}$.}
	\label{fig:mode decomposition}	
\end{figure*}

As described in the main text, we introduce two adimensional variables which are used for tuning the input conditions. Specifically, we normalized both the input width and offset to the FWHM $w_{\rm mode,\it\,n\rm=1}$ of the fundamental mode. Thus we define the two parameters $r_w$ and $r_s$ as follows: 
\begin{equation}
	r_{\rm w}=w/w_{\rm mode,\it\,n\rm=1}, 
\end{equation}
and 
\begin{equation}
	r_{\rm s}=x_{0}/w_{\rm mode,\it\,n\rm=1}.
\end{equation}
By varying $r_{\rm w}$ and $r_{\rm s}$, we can then scan the mode content of different input Gaussian beams. In Fig.~\ref{fig:mode decomposition}, we show how the input mode content varies with $r_s$ for four values of $r_w$. Specifically, here we consider up to $N=30$ modes, which are depicted on the top of the figure.	As Fig.~\ref{fig:mode decomposition} shows, when the width of a Gaussian beam coincides with that of the fundamental mode (i.e., $r_{\rm w}=1$), then $|C_1|^2=1$ and $|C_n|^2=0$ for $n>1$. To the contrary, as soon as either $r_{\rm w}$ or $r_{\rm s}$ are different from 1, the mode content becomes non trivial. 



Finally, the blue lines in Fig.~\ref{fig:mode decomposition} represent the value of the quantity $\sum_{n=1}^{N}|C_n|^2$ as a function of $r_{\rm s}$. As it can be seen, either shifting the beam with respect the fiber axis, or enlarging its waist produce a drop of the blue curve. This indicates that the mode truncation is too strict for allowing a proper mode decomposition. Indeed, beams whose associated values of $r_w$ or $r_s$ are too high cannot be fully decomposed by only 30 modes. 
Of course, increasing the number of computed modes boosts up the accuracy of the decomposition method, thus allowing to properly decompose beams with larger values of $r_w$ and $r_s$ at the expense of the computational time.
However, for the experimental conditions considered in this work, we found that considering 15 modes for decomposing the input beam was a good compromise to match the validity of the decomposition mode with the accuracy of our simulations (as well as with the reasonableness of the computational time).
%
%
%
%
%

\pagebreak
\subsection{Dependence of SSFS and group delay on the mode composition of multimode solitons}

\begin{figure*}[htbp]
		\centering
		\includegraphics[scale=1]{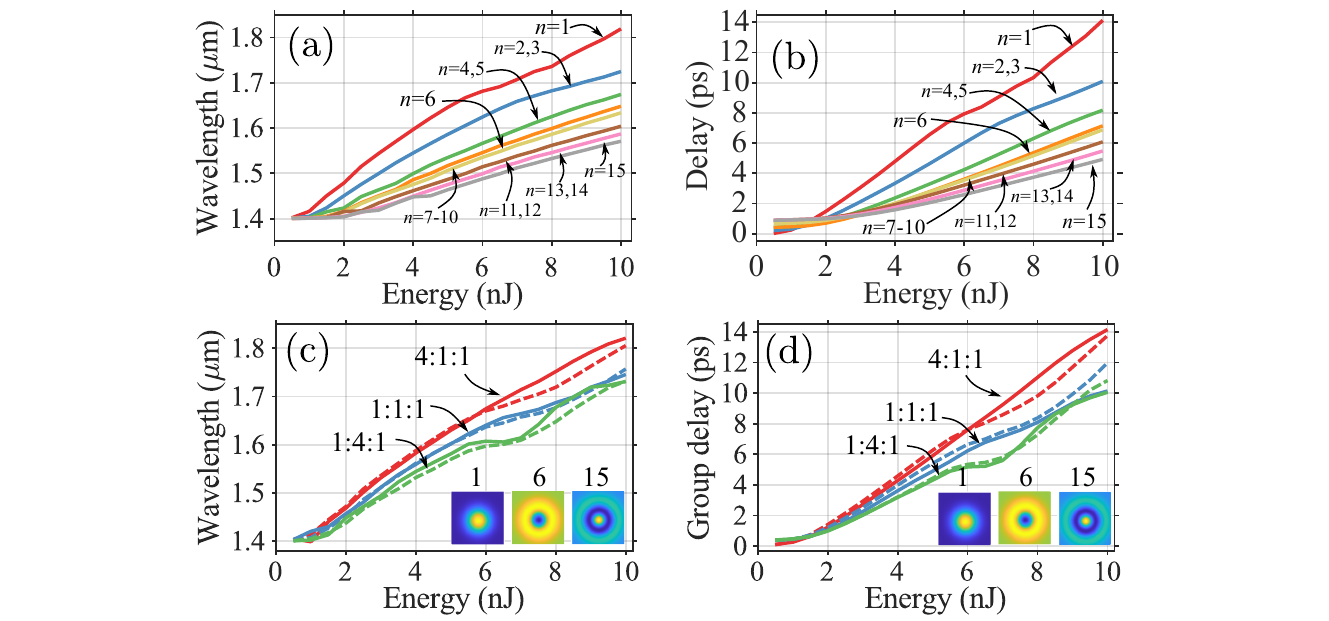}	
		\caption{
			(a,b) SSFS in (a) and GD in (b) of a single mode soliton as a function of input energy. 
			(c,d) SSFS in (c) and GD in (d) of multimode solitons as a function of input energy, plotted by solid lines. 
			The weighted SSFS and GD are plotted using dashed lines. The mode energy ratios of mode 1, 6, 15 are 4:1:1, 1:1:1, and 1:4:1, respectively. The fiber length is $\rm2\,m$.	
		}
		\label{fig:MMSs_SSFS_GD}	
	\end{figure*}

This section aims at investigating the dependence of SSFS and GD on the mode content of MMSs. In order to achieve this task, we keep all the conditions described in Sec.~2 of the main text. In the first instance, we consider the case of solitons which have a trivial mode composition, i.e., the consists of a single mode. The latter is not limited to the fundamental mode, but it varies from mode 1 up to mode 15. In Figs.~\ref{fig:MMSs_SSFS_GD}(a,b), we plot the SSFS and GD of such single mode solitons as a function of the input pulse energy. As it can be seen, the behavior is strongly affected by the mode content: both the SSFS and the GD are the largest when the soliton is carried by only the fundamental mode. The other way around, HOMs provide reduced effects. 

In the second instance, we consider the case of solitons which are made of only 3 modes. For sake of simplicity, we limit our analysis by considering only radial modes. Thus, the soliton is carried by only mode 1, 6, and 15. The latter are depicted in the inset of Fig.~\ref{fig:MMSs_SSFS_GD}(c,d). Here, we consider several cases, which only differ because of the ratio among the energy associated to each mode. In the first case, the fundamental mode is the most populated as its energy is fourfold higher than that of the remaining modes. In Fig.~\ref{fig:MMSs_SSFS_GD}(c,d), this case is indicated by the mode energy ratio, i.e., $4:1:1$. Then, we consider the case of equally populated modes (indicated by mode energy ratio  $1:1:1$). Finally, in the last case the 6-th mode is the most populated, being the mode energy ratio $1:4:1$. 
In Fig. \ref{fig:MMSs_SSFS_GD}(c,d), the solid lines represent the numerically calculated output SSFS and GD as a function of $E_{\rm in}$, respectively. Coherently with the results about single mode solitons, we found that MMSs carrying a larger portion of low order modes experience larger SSFS.

At last, we propose to analyse the features of MMSs by comparison with that of single mode solitons. In particular, we aim at comparing SSFS and GD of a MMS with the average values of that of single mode solitons. To do so, we consider a weighted average, in which the weight is given by the mode content of the MMS. Thus the averaged SSFS is calculated as $\overline{\lambda}_{\mathrm{SSFS}} = \sum_nP_{n}\lambda_{\mathrm{SSFS},\, n}/\sum_nP_{n}$, where $P_{n}$ is the peak power of mode $n$ of the MMS at the output and $\lambda_{\mathrm{SSFS},\, n}$ is the SSFS of the single mode soliton with mode $n$ in Fig.~\ref{fig:MMSs_SSFS_GD}(a). The resulting values are shown as dashed lines in Fig.~\ref{fig:MMSs_SSFS_GD}(c). 
An analogous weighted average can be applied for GD, which provide the dashed lines in Fig.~\ref{fig:MMSs_SSFS_GD}(d).
As it can be seen, the dashed lines are rather closer to the solid ones in both Fig.~\ref{fig:MMSs_SSFS_GD}(c) and (d) as long as the input energy does not exceed the limit value of 6 nJ, which is the threshold for the appearance of a second soliton as a result of the pulse fission. This indicates that MMSs can be seen as the overlap of single mode solitons, at least for what concerns SSFS and GD.

	
	\pagebreak
	\section{Group delay and SSFS without self-steepening.}
	
	%
	
\begin{figure*}[htbp]
		\centering
		\includegraphics[scale=1]{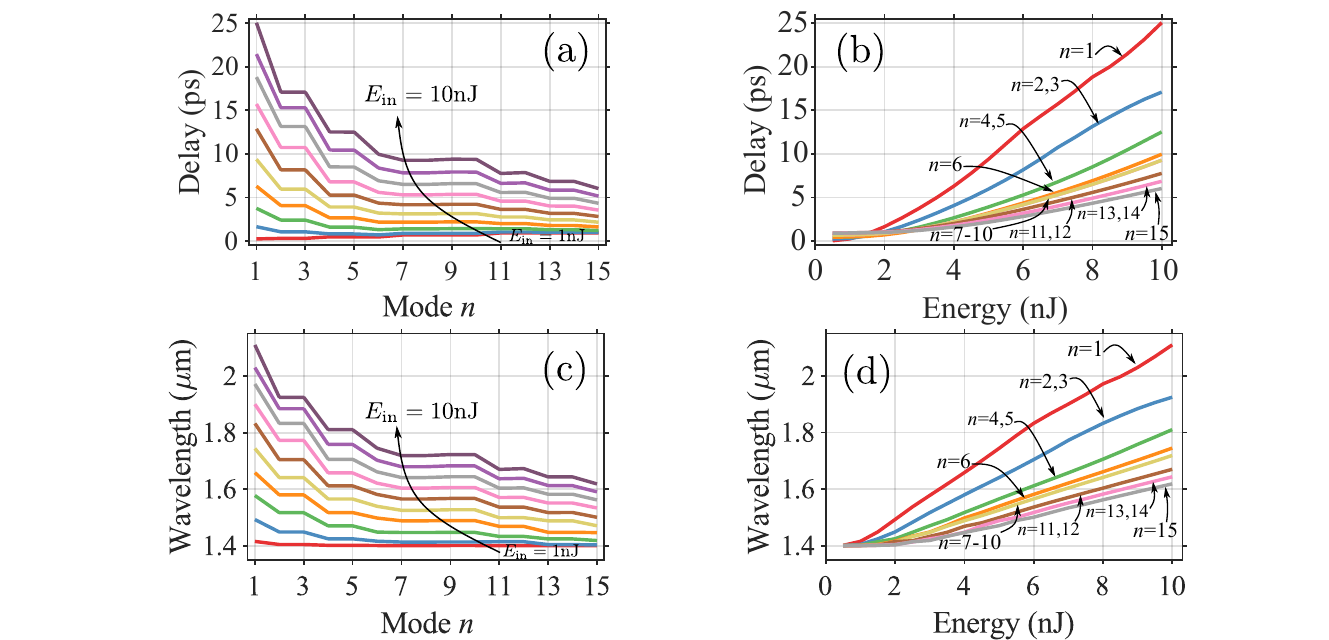}	
		\caption{Evolution of group delay and SSFS of a single mode soliton	in the absence of self-steepening.	(a,b) Group delays as a function of mode $n$ (a) and input energy (b). (c,d) SSFS vs. mode $n$ (c) and input energy (d).
		}
		\label{fig:SSFS_without_SS}	
	\end{figure*}

	It the main text, we have ascribed the soliton frequency shift and group delay to Raman effect. However, it is well known that self-steepening may produce similar influences on the pulse proprieties \cite{Agrawal2013}. Here, we prove the validity of our statement, by comparing simulations that are run in the absence of self-steepening, and compare them to the results reported in Fig.~1(c,d) in the main text and Fig.~\ref{fig:MMSs_SSFS_GD}(a,b). To do so, we run simulations with the same conditions of that reported in the main text, but removing the time derivative term in Eq.(3). The results are plotted in Fig.~\ref{fig:SSFS_without_SS}.
	
	By comparing results reported in Fig.~1(c,d) in the main text and Fig.~\ref{fig:MMSs_SSFS_GD}(a,b) with Fig.~\ref{fig:SSFS_without_SS}, one can notice that without self-steepening, both SSFS and group delay are larger with respect to the former cases. This indicates that self-steepening reduces the SSFS, which, on the other hand, is induced by Raman effect. We highlight that this evidence is in agreement with previous analytical studies on single mode fibers, which were reported in Ref.\cite{Voronin2008}.
	
\end{document}